# Polar domain walls trigger magnetoelectric coupling

*Josep Fontcuberta[a], Vassil Skumryev[a,b,c], Vladimir Laukhin[a,b], Xavier Granados[a] and Ekhard K. H. Salje[d]*

[a] Institut de Ciència de Materials de Barcelona (ICMAB-CSIC), Campus UAB, 08193, Bellaterra, Catalonia, Spain.

[b] Institució Catalana Recerca & Estudis Avançats, 08010 Barcelona, Catalonia, Spain.

[c] Univ. Autònoma de Barcelona, Dept. Física, 08193 Bellaterra, Catalonia, Spain,

[d] Department of Earth Sciences, University of Cambridge, Downing Street, Cambridge CB3 2EQ UK

*Interface physics in oxides heterostructures is pivotal in material's science. Domain walls (DWs) in ferroic systems are examples of naturally occurring interfaces, where order parameter of neighboring domains is modified and emerging properties may develop. Here we show that electric tuning of ferroelastic domain walls in $SrTiO_3$ leads to dramatic changes of the magnetic domain structure of a neighboring magnetic layer ($La_{1/2}Sr_{1/2}MnO_3$) epitaxially clamped on a $SrTiO_3$ substrate. We show that the properties of the magnetic layer are intimately connected to the existence of polar regions at twin boundaries of $SrTiO_3$, developing at T << 105 K, that can be electrically modulated. These findings illustrate that by exploiting the responsiveness of DWs nanoregions to external stimuli, even in absence of any domain contribution, prominent and adjustable macroscopic reactions of neighboring layers can be obtained. We conclude that polar DWs, known to exist in other materials, can be used to trigger tunable responses and may lead to new ways for the manipulation of interfacial emerging properties, such as 2DEGs.*

For many years strontium titanate SrTiO$_3$ has been the workbench where disruptive concepts and properties in solid state physics have been explored, discovered and tested[1]. Upon cooling, SrTiO$_3$ (STO) displays[2] a transition from a cubic phase (C) to a tetragonal (T), centrosymmetric phase at T$_{CT}$ ≈ 105 K. This structural transition involves octahedra rotations along all three cubic axis and it is accompanied by elastic strains[3, 4]. It was earlier recognized, that temperature-dependent permittivity of STO displayed Curie-like behavior down to ≈ 35 K signaling a strong tendency towards a ferroelectric instability. However, no ferroelectricity was observed and it was argued that this was blocked by quantum fluctuations, and STO is understood as a quantum paraelectric [5]. Associated with the C-T transition, ferroelastic twin domains (FTD) are formed. It has been found that upon cooling, the population of distinctly oriented FTD remains basically unperturbed[6-8]. However, application of electric fields or strain promotes reordering of FTD[7] and even the appearance of ferroelectric signatures[9, 10], including switchable ferroelectric Bloch lines[11-13], emerging at T ≤ 50 K. Moreover, as the T-phase results from a cooperative rotation of octahedra, regions with clockwise and anti-clockwise rotation should exist, producing antiphase boundaries (APB). Concomitantly, domain walls (DWs) are formed where local symmetry can be further lowered; the interplay of coexisting domains and DWs, both potentially responsive to external stimuli, challenges detailed understanding of the STO properties. The consensus achieved could be summarized as follows: i) bulk STO can be driven to be polar under suitably large electric field or mechanical stress[9, 10], ii) at ≈ 80 K, the local symmetry is reduced to a triclinic-like[11, 14-16], and iii) ferroelastic DWs display a polar response, incipient at ≈ 105 K[12], but clearly visible at T$_1^*$ (≈ 80 K), and further enhanced below T$_2^*$ (≈ 45 K)[6, 11].

STO is also the most used substrate for epitaxial growth of functional oxides of interest in spintronics, ferroelectric memories, etc. where it acts as an "inert" substrate and only epitaxial strain arising from structural mismatch with overgrown films, has been exploited to modulate film's properties[8, 17]. Imposed by the continuity of the metal-oxygen coordination polyhedral network, the elastic coupling across interface transmits any structural transformation of the substrate into the overgrown thin film[18]. Indeed, it has been shown that the magnetization of a La$_{1-x}$Sr$_x$MnO$_3$ (LSMO) thin film grown on STO(100) displays remarkable changes upon crossing T$_{CT}$[19-22], which were attributed to changes of the LSMO magnetic domain structure when shacked by the structural transition occurring in STO[19-21].

Here, we shall demonstrate that STO can be used as "active" substrate to modify the magnetic properties of a LSMO epitaxial overlayer, namely the magnetic domain configuration, by exploiting its various low-temperature structural transformations and the presence of polar nanoregions at DWs. The results here presented constitute a fresh example of making use of the polar nature of domain walls in one layer to manipulate the magnetic domains in a neighboring layer. This approach differs in its essence from the much investigated strain control - including through piezoelectric substrates - of the magnetic properties of LSMO and other oxides[23-25], where domains rather than DWs are the primarily responsive elements of the substrate. To enhance the sensitivity of the magnetic structure of the manganite layer to tiny structural changes, the LSMO composition has been chosen to be close to the verge of the ferromagnetic-antiferromagnetic phase transition; this is achieved in La$_{1/2}$Sr$_{1/2}$MnO$_3$[26, 27]; an instability further reinforced by the tensile strain imposed by the STO substrate[27]. Moreover, it will be shown that upon applying a suitable electric field to the STO substrate, an unparalleled response, and even a reversal of magnetic moment ($m$) of the LSMO layer can be obtained.

**Results**

**Oxide epitaxial heterostructures**

Several LSMO thin films of different thickness were grown on STO(001) single crystals. Data reported here correspond to films of 34 nm and 10 nm. The STO substrate induces a tensile strain (1.5 %). X-ray measurements have been done to assess the quality of the films, their strain state (Supplementary Information 1) and to estimate the film thickness[27]. A LSMO film (20 nm) was grown on LaAlO$_3$(001) single crystalline substrate under identical conditions, for comparison purposes [27].

**Magnetic properties of films**

The 34nm thick LSMO on STO(001) sample displays a Curie temperature and a saturation magnetization typical for strained films of this composition (Supplementary Information 2)[27]. The low temperature magnetic moment ($m$(T)) data recorded upon warming, either after magnetic field-cooling (H-FC) or after a zero-field-cooling (H-ZFC), do not merge (Figure 1(left-bottom) (circles)). This is a common feature of ferromagnetic systems when measured at small magnetic fields. Of more interest is the observation that the H-ZFC branch displays remarkable changes of slope at some temperatures: $T_0 \approx 108$ K, $T_1 \approx 90$ K and $T_2 \approx 45$ K. One immediately recognizes that $T_0$ closely coincides with the cubic-to-tetragonal transition temperature $T_{CT}$. This is a fingerprint of the magnetic domain reordering of LSMO when the STO substrate transforms to tetragonal phase and a complex array of crystalline domains is formed[19-22]. This magnetic anomaly is also visible in the H-FC branch at $T_{CT}$. These changes of slope in $m$(T) are indications of a change of the magnetic domain structure or its stiffness. As LSMO does not present any structural transformation or prominent magnetic anisotropy change in this temperature range, the observed anomalies of $m$(T) at $T_1$ and $T_2$ should be related to modification of the magnetic domain structure associated to changes in the STO substrate.

As emphasized by sketch in Figure 1(left-top), $T_1 \approx T_1^*$ ($\approx 90$ K) - where triclinic distortions, compatible with polar DW regions develop, and $T_2 \approx T_2^*$ ($\approx 45$ K) - where the polar character in STO is reinforced[6, 11]. For comparison, we include in Fig. 1(bottom) the $m$(T) data of a similar LSMO sample, 10 nm thick (triangles); the pronounced features at $T_{CT}$, $T_1$ and $T_2$ can also be appreciated. These data indicate that the magnetic state of LSMO is a probe to the structural, microstructural and dielectric changes occurring at the substrate. As FTD orientation can be modified by electric fields[7] and polar (piezoelectric) DWs sensitive to electric fields, are formed at temperatures below $T_{CT}$[11, 12], it should be expected that application of a suitable electric field to the STO substrate would induce a domain and domain wall reconstruction, which in turn should impact, via interface magnetoelastic coupling, the magnetic state of the LSMO layer (Fig. 1 right).

**Electrical biasing of SrTiO$_3$ substrate. Magnetic properties of LSMO films**

To demonstrate this effect, thermomagnetic curves were measured at a fixed magnetic field (H$_{meas}$) under different constant electrical field (E) (Figure 2a) (Supplementary

Information 3). Prior each run, the sample was cooled down to 5 K from T >> $T_{CT}$, in an E-field identical to the E-field under which the corresponding thermomagnetic curve was recorded. Any possible E-field effect on the magnetic response should be more pronounced close to the magnetic coercivity field (≈60-70 Oe, for this particular sample) . This is why in the experiment of Figure 2b, the sample was cooled down in a negative magnetic field (-65 Oe) and $m$(T) was measured either during the cooling process or using a positive $H_{meas}$ = 65 Oe field upon warming. The comparison of data recorded upon warming (Fig. 2b-top), under V = 0, +150 V, and +210 V shows that whereas for T > 80 K, the $m$(T) data match those recorded at V = 0 V, this is not the case for T < 80 K where a strong dependence of $m$(T) on V is observed. We note that the magnetic moment measured (after the H-ZFC procedure) is smaller when $m$(T) measurements are performed under E-field. Further after, the $m$(T) data recorded during the H-FC process (Fig. 2b-bottom) ($H_{meas}$ = - 65 Oe) also displays the characteristic feature at $T_{CT}$ ≈ 108 K, and splitting between the E-ZFC and E-FC. Consistently, $|m|$ is smaller when measured under E-field (Fig.2b), implying electric-field induced magnetic "hardening".

Data in Figures 2b indicate that an electric-field-induced modification of the crystalline domain structure of STO translates into different magnetic domain structures in the LSMO layer and therefore into magnetization differences. In this scenario, it can be expected that similar effect should be visible in isothermal measurements. We show in Figure 3a isothermal (5 K) measurements of $m$ as a function of the applied E-field. Prior to measurements, sample was first cooled down to 5 K under V = 0 V and nominally H = 0 Oe conditions. The measuring magnetic field $H_{meas}$ = 55 Oe was applied at 5 K and $m$ was recorded subsequently. First, a sequence of measurements was performed for $t$ ≈ 50 min to monitor the eventual presence of E-independent magnetic aftereffects[28]. As shown in Fig. 3a, the magnetic moment only varies by about 2 % in 50 min. Next, an E-field step sequence (Fig. 3 (inset)) was applied as indicated. It follows from data in Fig. 3a that any V change, irrespectively of its sign, induces an increase of magnetic moment. This implies that the magnetization is driven towards its equilibrium value by V steps (Supplementary Information 4). Therefore, it should be possible to reverse the magnetic moment of the sample by a V-step, if the measurement is performed under appropriate H > 0 on a sample previously cooled down in H < 0. Indeed, this is clearly observed in Fig. 3b where $m$ is recorded isothermally (21 K) after cooling in H = -65 Oe and measuring in +55 Oe.

The experiments described so far signal the active role of STO and its responsivity to electric field stimulus to induce changes in the magnetic state of the LSMO layer. The rationale behind is directly related to the genuine properties of the low-temperature phase of STO, namely the presence of polar regions with their concomitant piezoresponse (Fig. 1 right).. It thus follows that if similar experiments were performed using substrates where this instability was absent, one should not expect any change in the magnetic state of the magnetic layer upon applying a voltage at the substrate. We have performed such experiments on a LSMO (20 nm) film grown on $LaAlO_3$(001) (Supplementary Information 2). In Figure 3c we show the magnetic moment measured at 5 K, after cooling in nominally H = 0, under $H_{meas}$= 230 Oe (a value similar to the coercivity field) when successive V-steps are applied. It is obvious that the applied electric field does not modify the magnetic moment of the LSMO layer.

## Discussion

The results presented above show that the magnetic moment of the LSMO films, when measured in absence of electric fields (Fig. 1), displays remarkable changes at some characteristic temperatures that closely coincide with: (i) cubic-to-tetragonal structural transition, (ii) the onset of the reported lowering of symmetry to a triclinic-like phase[14] where polar domain walls are formed[11], and (iii) at some lower temperature, where a polar nature of the DW appears to be reinforced[11]. These effects connect the changes in the domain structure and DW properties[11, 29] of the substrate with those of the magnetic films, epitaxially clamped on its surface. In fact, coupling between ferroic domain structure in piezo substrates and overgrown films has been shown to be a suitable way to modify the magnetic properties of the latter by applying an electric field to the former[30, 31]. The magnetic layer is thus a sensitive probe of the substrate structural transformations.

More remarkable is the observation that when cooling the sample from T >> $T_{CT}$ under an E-field, the magnetic structure of the film is modified and thus its magnetic response changes accordingly. This observation reflects that the microstructure of the STO substrate, namely its FTD distribution is modified, during the cooling process, by the presence of E field. At this point a question arises: is the observed effect related to the already reported redistribution of twin domains due to their anisotropic electrostriction[7]? To address this point we note that we observe E-field effects developing at T < 90 K (Fig. 2b), but in this temperature region, and within the sensitivity of X-ray diffraction experiments, effects were only perceptible when E > 8 kV/cm. Therefore, as our applied electric field (Supplementary Information 3) is estimated to be ≈ 2.1kV/cm, this mechanism is unlikely to be prominent. Instead, the polar nature of nanoregions in STO, either within the domains or at the domain walls, provides a simple framework to explain the observed effects. Indeed, in the temperature-dependent $m$(T) experiments (Fig. 2), we have observed that substantial E-field effects emerge below ≈ 90 K, close to the temperature where polar domain walls have been already identified[11]. Therefore, the applied E-field could be effective in changing the twin domain distribution in STO and correspondingly sensed by the LSMO magnetization, as observed. Consistently, an even larger effect is observed at T below ≈ 45 K, which is where the DWs in STO appear to reinforce their polar character and their mobility[11]. It is interesting to note that the applied E-field is mainly concentrated under the electrodes (Figs. 2a and S2), thus affecting only about 20% of the LSMO volume. Still the E-induced changes of |m| are remarkably large (up to 50 %, see Supplementary Information 4). This indicates that dielectric domain modification in the substrate promotes magnetic avalanches in the LSMO, as commonly found in many magnetic systems[32].

In brief, an applied electric field, benefiting from the presence of polar nanoregions in STO and its associated piezoresponse, produce dramatic changes in the magnetic state of an overlying layer, magnetoelastically coupled to the substrate. In an appropriate magnetic state, the magnetic moment of the films can be even reversed by an electric field applied to its substrate. These results show, for the first time, that STO can be used as active substrate and they suggest new ways to explore and exploit its properties. We strength that STO domain wall responsiveness to electric fields, due to its inherent polar nature, is at the heart of the observed response which thus constitute an example of exploiting polar domain walls to trigger a magnetic response. The experiments here reported have been made in the simplest back-contacting configuration, which although good enough to demonstrate the soaked effects, can be largely improved by suitable nanotechnology contacting procedures[33]. The impact of these findings can be far reaching. For instance, in the celebrated 2DEGs formed at LAO-STO (001) and (110) interfaces[34], and particularly in gating experiments[35, 36] where large electric fields are applied across films, the piezoresponse shown here could have interesting impacts on the observed properties of the gas and their anisotropy. Finally, it is worth noticing that polar DWs are not unique to STO but other systems have been shown to have ferroelectric DWs, even offering the possibility to be operated at room temperature[37]. We thus conclude that our results constitute a contribution towards domain wall nanoelectronic[38].

## Methods

### Thin film growth and structural characterization

Several La$_{1/2}$Sr$_{1/2}$MnO$_3$ (LSMO) thin films of different thickness were grown on as-received STO(001) single crystals (5 x 5 mm$^2$) by Pulsed Laser Deposition. Films have been deposited at 725°C in 0.2 mbar oxygen pressure with subsequent free cooling in 100 mbar oxygen pressure. The STO substrate induces a tensile strain (1.5 %). X-ray measurements have been done to assess the quality of the films, their strain state (Supplementary Information 1) and to estimate the film thickness[27]. For comparison purposes, a LSMO film (20 nm) was grown on LaAlO$_3$(001) single crystalline substrate under identical conditions. The growth rate has been calibrated by measuring, by X-ray reflectivity (XRR), the thickness of some on-purpose prepared films. The X-ray diffraction (XRD) and XRR measurements were carried out using Cu-Kα radiation, in a Siemens D-5000 diffractometer and a Rigaku RU-200B diffractometers. Reciprocal space maps (RSM) were collected using a Bruker 1T8 Advance diffractometer equipped with a bidimensional detector, and used to determine in-plane cell parameters.

### Magnetic measurements

Magnetic measurements were conducted using a SQUID magnetometer (QD) equipped with an in-house made wiring set-up that allows applying in-situ electric fields on samples. In all experiments reported here, the magnetic field has been applied in plane of the film. It turned out that all the fine details of the sample magnetic moment variations are very sensitive to the specific measuring conditions, i.e. the strength of the magnetic field in which the sample is cooled down and the strength of the magnetic field, in which the magnetic moment is recorded as well as the temperature from which the sample is cooled down, and the number of measurements/cycles (magnetic history). In magnetic zero-field cooling (H-ZFC) experiments the samples are cooled down from temperatures well above $T_{CT}$, in nominally zero magnetic field after carefully demagnetizing the Mu-metal shield surrounding the SQUID magnet and also quenching the magnet to get rid of the trapped field.

### Electric biasing

For electric biasing, two wires were connected to the back of the crystalline substrate, along the edges of the substrate using conducting graphite paste (Fig. 2a). The distance between the wires is of about 4 mm. Electric field was applied by voltage V (up to ± 210 V) using a Keithley 2400 sourcemeter to the connecting wires. Due to the geometry used in the experiment, the electric field inside the dielectric STO is inhomogeneous: mostly concentrated under the electrodes with a relatively large value ($E \approx V/d$, where d is the substrate thickness), reaching up to 2.1 kV/cm for the maximum voltage used), and much smaller field E' between the electrodes; for the used geometry, at the midline between electrodes, E' averages to ≈ 0.1 kV/cm (for V = 210 V) and depends on the permittivity of STO (Supplementary information 3). When appropriate, we use E as a label of the applied electric field and accordingly, the applied fields are within the range: -2.1 kV/cm < E < +2.1 kV/cm. During measurements, the current flowing was monitored and it was always found to be limited by the sensitivity of the test set-up (<10 nA); no changes were observed either when applying 210 V at the electrodes or upon varying the temperature, confirming that no appreciable current was flowing across the device and confirming the absence of shortcuts. Magnetic measurements were typically started within 20-30 sec after setting the voltage values. It is worth mentioning that, when exploring the $m$(T,V) data under different V-bias, such as in Figure 2b-top, results depend slightly on sample history.

## Acknowledgments

This work was supported by the Spanish MAT2011-29269-C03 and MAT2014-56063-C2-1-R projects and the Generalitat de Catalunya (2014 SGR 734) project. We are thankful to F. Sánchez, G. Radaelli and D. Gutiérrez for growing the films and making the preliminary characterization of the films.

## Author contributions

J.F. designed and conceived the experiments. The experimental set-up and samples contacting were done by V.L. V.S. performed the magnetic measurements and X. G. did the E-field calculations. Data analysis was done by J.F, V.S., V.L. and E.K.H.S. J.F. wrote the paper. All authors discussed data and commented on the paper.

## Additional Information

## References

[1] J. G. Bednorz and K. A. Müller, Perovskite-Type Oxides the new approach to High-Tc Superconductivity, *Nobel lecture* December 8 (1987).

[2] E. K. H. Salje, M. C. Gallardo, J. Jimenez, F. J. Romero, and J. del Cerro, The cubic-tetragonal phase transition in strontium titanate: excess specific heat measurements and evidence for a near-tricritical, mean field type transition mechanism, *Journal of Physics-Condensed Matter* 10, 5535 (1998).

[3] J. Slonczew and H. Thomas, Interaction of Elastic Strain with Structural Transition of Strontium Titanate, *Physical Review B* 1, 3599 (1970).

[4] P. A. Fleury, J. F. Scott, and J. M. Worlock, Soft Phonon Modes and 110 degrees K phase Transition in $SrTiO_3$, *Physical Review Letters* 21, 16 (1968).

[5] K. A. Muller and H. Burkard, $SrTiO_3$ - Intrinsic Quantum Para-Electric Below 4-K, *Physical Review B* 19, 3593 (1979).

[6] A. V. Kityk et al. Low-frequency superelasticity and nonlinear elastic behavior of $SrTiO_3$ crystals, *Physical Review B* 61, 946 (2000).

[7] J. Sidoruk et al., Quantitative determination of domain distribution in $SrTiO_3$-competing effects of applied electric field and mechanical stress, *Journal of Physics-Condensed Matter* 22, 235903 (2010).

[8] A. Buckley, J. P. Rivera, and E. K. H. Salje, Twin structures in tetragonal $SrTiO_3$: The ferroelastic phase transition and the formation of needle domains, *Journal of Applied Physics* 86, 1653 (1999).

[9] J. Hemberger, P. Lunkenheimer, R. Viana, R. Bohmer, and A. Loidl, Electric-Field-Dependent Dielectric-Constant and Nonlinear Susceptibility in $SrTiO_3$, *Physical Review B* 52, 13159 (1995).

[10] J. Hemberger et al., Quantum paraelectric and induced ferroelectric states in $SrTiO_3$, *Journal of Physics-Condensed Matter* 8, 4673 (1996).

[11] E. K. H. Salje, O. Aktas, M. A. Carpenter, V. V. Laguta, and J. F. Scott, Domains within Domains and Walls within Walls: Evidence for Polar Domains in Cryogenic $SrTiO_3$, *Physical Review Letters* 111, 247603 (2013).

[12] P. Zubko, G. Catalan, A. Buckley, P. R. L. Welche, and J. F. Scott, Strain-gradient-induced polarization in $SrTiO_3$ single crystals, *Physical Review Letters* 99, 167601 (2007).

[13] E. K. H. Salje and J. F. Scott, Ferroelectric Bloch-line switching: A paradigm for memory devices?, *Applied Physics Letters* 105, 252904 (2014).

[14] R. Blinc, B. Zalar, V. V. Laguta, and M. Itoh, Order-disorder component in the phase transition mechanism of O-18 enriched strontium titanate, *Physical Review Letters* 94, 147601 (2005).

[15] V. V. Laguta, R. Blinc, M. Itoh, J. Seliger, and B. Zalar, Sr-87 NMR of phase transitions in (SrTiO3)-O-16 and (SrTiO3)-O-18, *Physical Review B* 72, 214117 (2005).

[16] J. F. Scott, J. Bryson, M. A. Carpenter, J. Herrero-Albillos, and M. Itoh, Elastic and Anelastic Properties of Ferroelectric (SrTiO3)-O-18 in the kHz-MHz Regime, *Physical Review Letters* 106, 105502 (2011).

[17] F. M. Granozio, G. Koster, and G. Rijnders, Functional oxide interfaces, *Mrs Bulletin* 38, 1017 (2013).

[18] J. M. Rondinelli, S. J. May, and J. W. Freeland, Control of octahedral connectivity in perovskite oxide heterostructures: An emerging route to multifunctional materials discovery, *MRS Bulletin* 37, 261 (2012).

[19] D. Pesquera, V. Skumryev, F. Sanchez, G. Herranz, and J. Fontcuberta, Magnetoelastic coupling in $La_{2/3}Sr_{1/3}MnO_3$ thin films on $SrTiO_3$, *Physical Review B* 84, 184412 (2011).

[20] V. K. Vlasko-Vlasov et al., Direct magneto-optical observation of a structural phase transition in thin films of manganites, *Physical Review Letters* 84, 2239 (2000).

[21] V. K. Vlasko-Vlasov et al., Observation of the structural phase transition in manganite films by magneto-optical imaging, *Journal of Applied Physics* 87, 5828 (2000).

[22] Y. Segal et al. , Dynamic Evanescent Phonon Coupling Across the $La_{1-x}Sr_xMnO_3$/$SrTiO_3$ Interface, *Physical Review Letters* 107, 105501 (2011).

[23] A. Herklotz et al., Strain response of magnetic order in perovskite-type oxide films, *Philosophical Transactions of the Royal Society a-Mathematical Physical and Engineering Sciences* 372, 20120441 (2014).

[24] W. Eerenstein, M. Wiora, J. L. Prieto, J. F. Scott, and N. D. Mathur, Giant sharp and persistent converse magnetoelectric effects in multiferroic epitaxial heterostructures, *Nature Materials* 6, 348 (2007).

[25] Q. He et al., Electrically controllable spontaneous magnetism in nanoscale mixed phase multiferroics, *Nature Communications* 2, 225 (2011).

[26] A. Urushibara et al., Insulator-Metal Transition and Giant Magnetoresistance in $La_{1-x}Sr_xMNO_3$, *Physical Review B* 51, 14103 (1995).

[27] D. Gutierrez, G. Radaelli, F. Sanchez, R. Bertacco, and J. Fontcuberta, Bandwidth-limited control of orbital and magnetic orders in half-doped manganites by epitaxial strain, *Physical Review B* 89, 075107 (2014).

[28] S. Chikazumi, Physics of Ferromagnetism, *Oxford University Press* (1997).

[29] T. Zykova-Timan and E. K. H. Salje, Highly mobile vortex structures inside polar twin boundaries in $SrTiO_3$, *Applied Physics Letters* 104, 082907 (2014).

[30] V. Skumryev et al., Magnetization Reversal by Electric-Field Decoupling of Magnetic and Ferroelectric Domain Walls in Multiferroic-Based Heterostructures, *Physical Review Letters* 106, 057206 (2011).

[31] V. Laukhin et al., Electric-field control of exchange bias in multiferroic epitaxial heterostructures, *Physical Review Letters* 97, 227201 (2006).

[32] S. Papanikolaou et al., Universality beyond power laws and the average avalanche shape, *Nature Physics* 7, 316 (2011).

[33] L. J. McGilly, P. Yudin, L. Feigl, A. K. Tagantsev, and N. Setter, Controlling domain wall motion in ferroelectric thin films, *Nature Nanotechnology* 10, 145 (2015).

[34] A. Ohtomo and H. Y. Hwang, A high-mobility electron gas at the $LaAlO_3$/$SrTiO_3$ heterointerface, *Nature* 427, 423 (2004).

[35] A. D. Caviglia et al., Electric field control of the $LaAlO_3$/$SrTiO_3$ interface ground state, *Nature* 456, 624 (2008).

[36] G. Herranz et al., Engineering two-dimensional superconductivity and Rashba spin-orbit coupling in $LaAlO_3$/$SrTiO_3$ quantum wells by selective orbital occupancy, *Nature Communications* 6, 6028 (2015).

[37] S. Van Aert et al. , Direct Observation of Ferrielectricity at Ferroelastic Domain Boundaries in $CaTiO_3$ by Electron Microscopy, *Advanced Materials* 24, 523 (2012).

[38] G. Catalan, J. Seidel, R. Ramesh, and J. F. Scott, Domain wall nanoelectronics, *Reviews of Modern Physics* 84, 119 (2012).

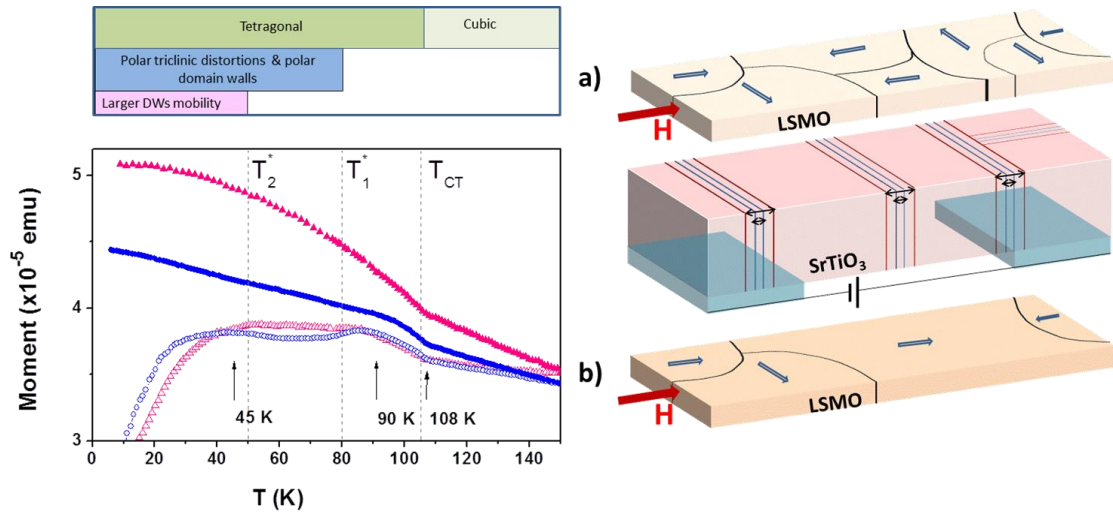

**Figure 1 | Phase diagram of SrTiO₃, magnetic moment of a LSMO/STO(001) film and coupling of magnetic domains by polar domain walls in STO.** *Left top*: representation of the known phase diagram of STO. *Left bottom*: temperature dependence of the magnetic moment of LSMO films on STO(001), of thickness 34 nm and 10 nm (blue circles and red triangles, respectively), measured on warming, after cooling in zero-magnetic field from room-temperature (△, ○ ), and on cooling (▲, ● ). Data have been collected using $H_{meas}$ = 60 Oe. For clarity, data for 34 nm and 10 nm samples have been shifted along the *m*-axis, to coincide, on warming, at $T_{CT}$. Vertical dashed lines indicate the temperatures ($T_{CT}$, $T_1^*$, and $T_2^*$) where the known transformations in STO occur. *Right central panel*: polar domain walls in SrTiO₃ whose width (horizontal double ended arrows) is modified by E-biasing. *Right panel* a) illustrates the ferromagnetic domain distribution in the LSMO film, pinned by STO domain walls, under a (small) measuring magnetic field (H). *Right panel* b) illustrates the domain configuration in LSMO film, under a (small) measuring magnetic field (H), when the pining strength provided by STO domain walls is reduced under E-bias.

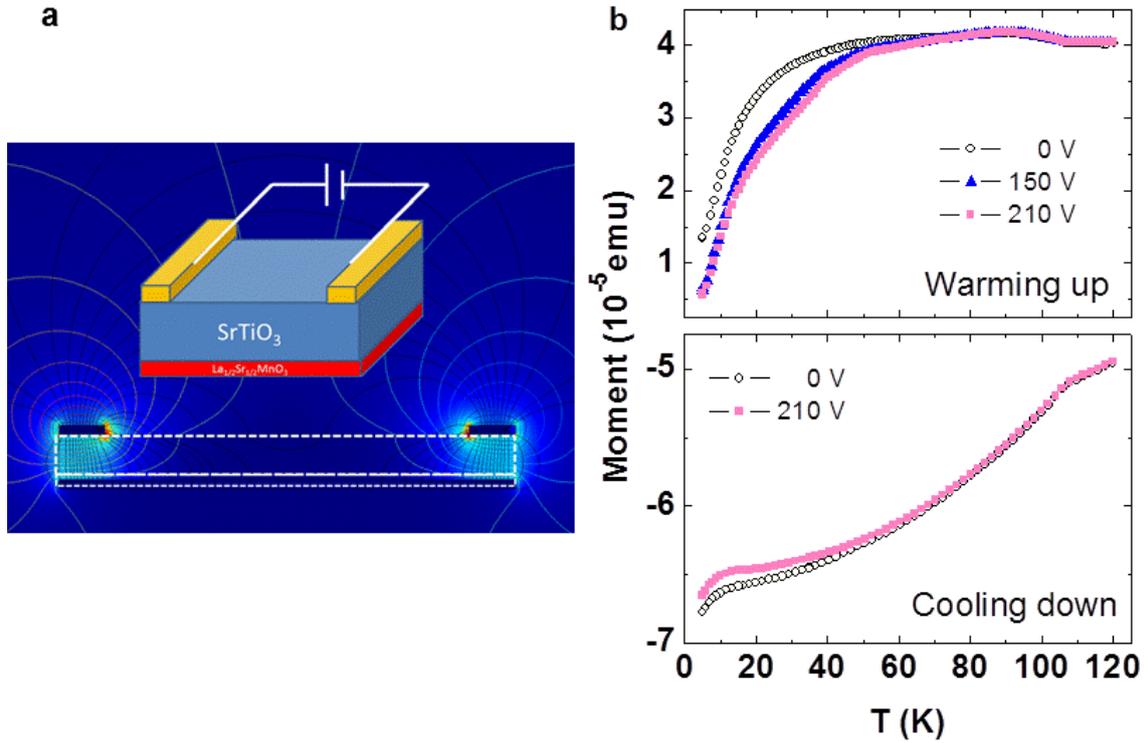

**Figure 2 | Electric connection of the sample and its electric-field and temperature-dependent magnetic moment. a,** Sketch of sample connection for electrical biasing and electric field distribution (Supplementary Information 3). Calculation has been made by using the finite elements solver COMSOL 4.4 package. Color indicates the strength of the electric field. **b,** Temperature dependence of magnetic moment of a LSMO (34 nm) sample measured at different electric fields. **Top**: on warming, under $H_{meas}$ = + 65 Oe, after cooling at different electric fields. During cooling and warming the electric field is identical. **Bottom:** on cooling down ($H_{meas}$ = -65 Oe), with and without electrical field applied. Prior each measurement the sample was cooled down from T >> $T_{CT}$ in magnetic field of -65 Oe.

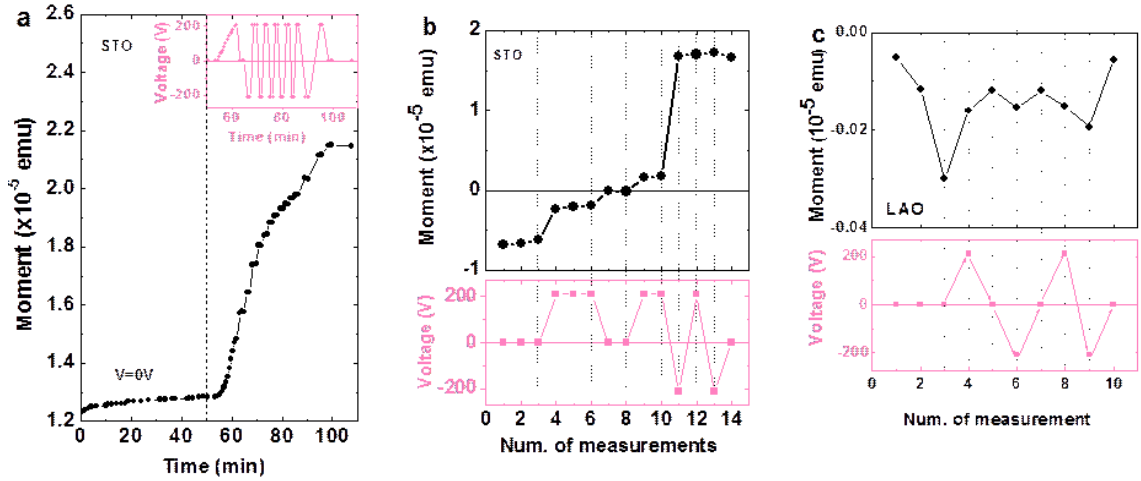

**Figure 3 | Isothermal changes of magnetic moment of LSMO films under E-field biasing.** Data for: LSMO (34 nm) on STO (**a**, **b**) and (20 nm) on LAO (**c**) are shown. **a**, at T = 5 K after nominally zero-field cooling and under $H_{meas}$ = 55 Oe; Inset: voltage steps applied to the sample during magnetic measurements. **b**, (top panel) at T = 21 K, after field cooling in H = -65 Oe, and at $H_{meas}$= 55 Oe; bottom panel, indicates the voltage steps applied during measurements. **c**, sample LSMO(20 nm) on (001)LAO at T = 5 K, after field cooling in H= -65 Oe and at $H_{meas}$= 230 Oe; bottom panel, indicates the voltage steps applied during measurements. The $H_{meas}$ values were chosen to be close to coercivity field of the corresponding sample. In all cases samples were cooled under V = 0. In all figures the connecting lines are guide to eyes.

**Supplementary Information**

# Polar domain walls trigger magnetoelectric coupling

*Josep Fontcuberta[a], Vassil Skumryev[a,b,c], Vladimir Laukhin[a,b], Xavier Granados[a]*
*and*
*Ekhard K. H. Salje[d]*

[a] Institut de Ciència de Materials de Barcelona (ICMAB-CSIC),
Campus UAB, 08193, Bellaterra, Catalonia, Spain.

[b] Institució Catalana Recerca & Estudis Avançats, 08010 Barcelona, Catalonia, Spain.

[c] Univ. Autònoma de Barcelona, Dept. Física, 08193 Bellaterra, Catalonia, Spain,

[d] Department of Earth Sciences, University of Cambridge, Downing Street, Cambridge
CB3 2EQ UK

**Supplementary Information 1**

**Epitaxial strain on La$_{1/2}$Sr$_{1/2}$MnO$_3$ films on (001)STO and (001)LAO.**
In Figure S1, reciprocal space maps of LSMO films of 20 nm, grown on STO(001) (a) and
LAO(001) (b) substrates, recorded around the corresponding (113) reflections of the
substrates, are shown.

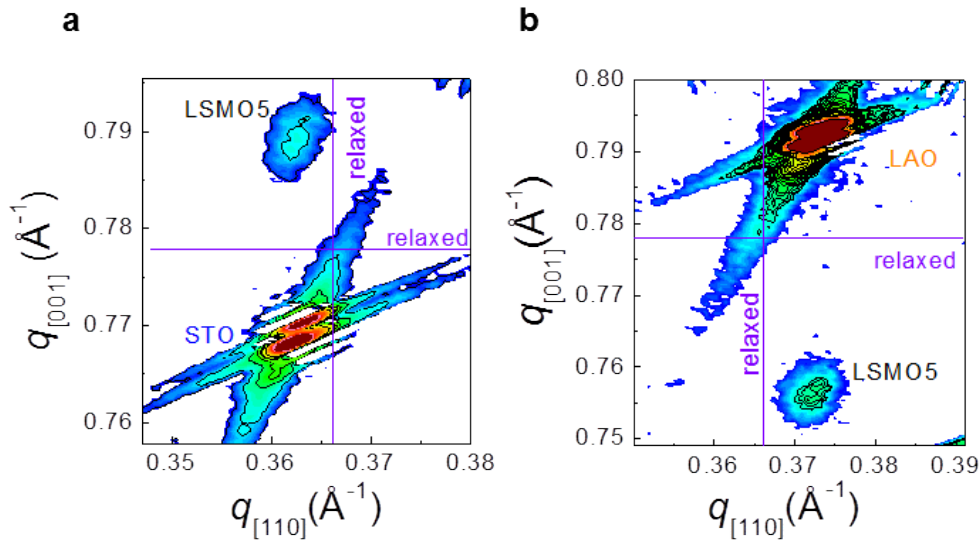

**Figure S1. Reciprocal space maps of LSMO films**. Around the reflection (113) of (a) STO and (b)
LAO. The tensile and compressive strain state of the LSMO on (001) STO and LSMO on (001)
LAO films are evidenced by the shift towards smaller (larger) *q*-values of the in-plane direction
of reflections of LSMO when compared to the bulk value (vertical line).

**Supplementary Information 2**

**Temperature dependence of the magnetization.**

In Figure S2 a) we show the temperature dependence of the magnetization (M) of the LSMO(34 nm)//(001)STO film, measured using H-ZFC and H-FC protocols as indicated ($H_{meas}$ = 100 Oe). It can be observed that the Curie temperature is of about 280 K. The characteristic changes of magnetization at the $T_{CT}$ of STO can be appreciated in both measurements. Measurements have also been performed at larger magnetic field (500 Oe and 1000 Oe). In all cases the kink of the M(T) data at $T_{CT}$ can be well appreciated although, as expected, the difference between H-ZFC and H-FC curves gradually vanishes when increasing the measuring field. In Figure S2 b) we show the corresponding measurement on the LSMO(20 nm)/LAO sample. No visible changes can be appreciated around 105 K. In agreement with earlier reports, the magnetization of LSMO films on LAO is smaller and films are magnetically harder.

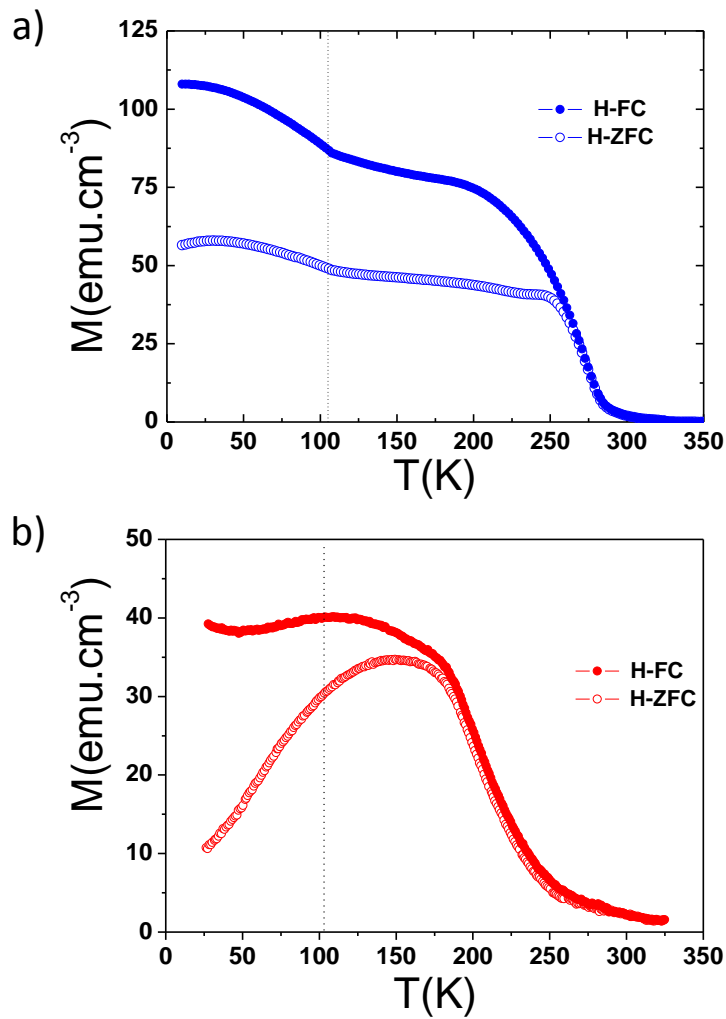

**Figure S2. Temperature dependence of the magnetization of a) LSMO(34 nm)/(001)STO and b ) LSMO(20 nm)/LAO film**. Measurements performed under $H_{meas}$ = 100 Oe, after cooling the sample either under magnetic field (H-FC) (solid symbols) or in nominally zero magnetic field (H-ZFC) (open symbols).

**Supplementary Information 3**

**Electric field distribution in the sample.**

The electric field at the sample and around was calculated by using the finite-elements solver COMSOL Multiphysics® Modeling Software package 4.4, at the device dimensions: a) substrate thickness 0.5 mm, b) lateral dimensions 5 x 5 mm², c) contact width 0.5 mm; d) contact length 5 mm that were placed on the crystal surface, opposite to LSMO face, along opposite edges of the substrate.

The calculated electric field (modulus) distribution is shown in Fig. S3. The color barcode indicates the strength of the electric field (V/m) at any point.

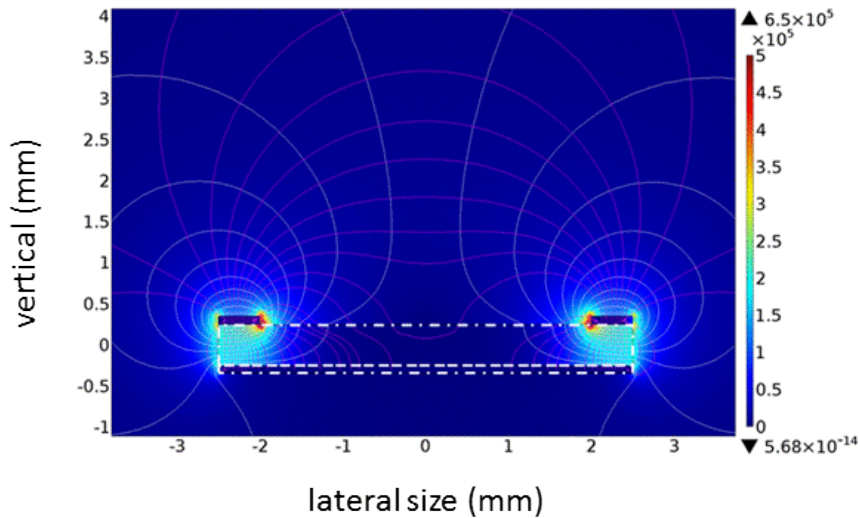

**Figure S3. Electric field distribution in the device**. Field units in the code bar (right vertical scale) are in V/m. Dimensions are in mm.

## Supplementary Information 4

## Temperature dependence of the magnetic moment.

In Figure S4 we show the m(T) data for the 34 nm LSMO sample recorded upon warming under $H_{meas}$ = 65 Oe, after an E-FC (V = +210 V) process; at some fixed temperatures during the warming process, the E-field was varied as indicated. For comparison we also include $m$(T) data collected upon warming under V = 0 and $H_{meas}$ = 65 Oe after a E-ZFC process. It can be observed that at the lowest temperatures, the magnetic moment recorded under E ≠ 0 is smaller than the corresponding value recorded under E = 0 This is in agreement with data in Figure 2b (main text). Next, we observe that at 22 K, when the E-field is zeroed, there is a rapid increase of magnetic moment, approaching - but still smaller- the $m$(T, E = 0) data (open symbols). This observation nicely illustrates that E-field steps, irrespectively of its sign, induce changes in the STO affecting the magnetic state of the LSMO, which gradually approaches its equilibrium state.

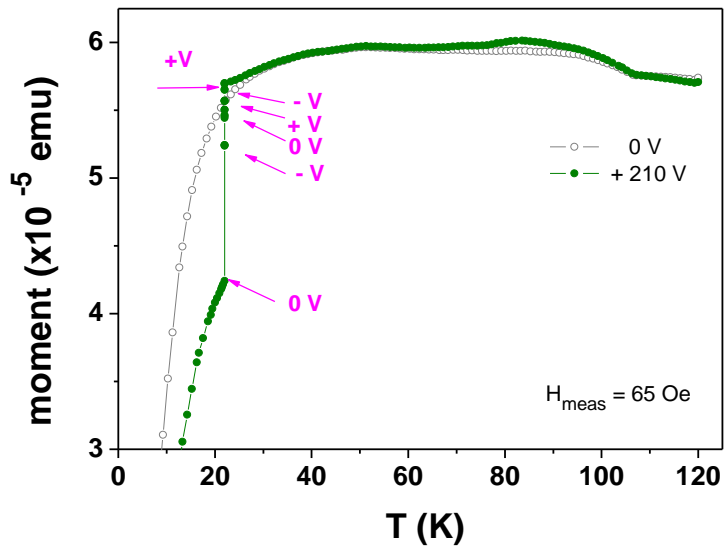

**Figure S4 | Electric field effects on the magnetic moment at fixed temperature and magnetic field on the LSMO film, on heating.** Temperature dependence of the magnetic moment of the 34 nm thick, LSMO film under $H_{meas}$ = 65 Oe. Prior each measurement the sample was cooled down from T >> $T_{CT}$ to 5 K, in magnetic field of -65 Oe, either without E-field applied (open symbols) or under electric field (210 V, solid symbols) and measurements were done upon warming. On sample cooled under E-field (solid symbols), arrows in the warming cycle indicate the V steps performed. First V changed was performed, at 22K, where V was zeroed. Subsequent V changes are indicated.